\documentclass[aps,pra,twocolumn,groupedaddress]{revtex4-2}

\usepackage{outlines}
\usepackage{dsfont}
\usepackage{enumitem}
\usepackage{braket}
\usepackage{amsmath}
\usepackage{mathtools}
\usepackage{amsfonts,amssymb,amsbsy}
\usepackage{amssymb}
\usepackage[T1]{fontenc}
\usepackage[ttdefault=true]{AnonymousPro}
\usepackage{graphicx}
\usepackage{MnSymbol}
\usepackage{xcolor}
\usepackage{gensymb}
\usepackage{xfrac}

\usepackage{hyperref}
\hypersetup{
	colorlinks=true,
	linkcolor=blue,
	filecolor=magenta,
	urlcolor=cyan,
}

\begin{document}


\title{Suppression of mid-circuit measurement crosstalk errors with micromotion}


\author{J. P Gaebler}
\thanks{Please direct any experimental questions to this author}
\email[]{John.Gaebler@Quantinuum.com}
\author{C. H. Baldwin}
\thanks{Please direct any benchmarking questions to this author}
\email[]{Charles.Baldwin@Quantinuum.com}
\author{S. A. Moses}
\author{J. M. Dreiling}
\author{C. Figgatt}
\author{M. Foss-Feig}
\author{D. Hayes}
\author{J. M. Pino}

\affiliation{Honeywell Quantum Solutions \\ 303 S. Technology Ct, Broomfield, Colorado 80021, USA}

\date{\today}

\begin{abstract}
Mid-circuit measurement and reset are crucial primitives in quantum computation, but such operations require strong interactions with selected qubits while maintaining isolation of neighboring qubits, which is a significant challenge in many systems. For trapped ion systems, measurement is performed with laser-induced fluorescence. Stray light from the detection beam and fluorescence from the measured ions can be significant sources of decoherence for unmeasured qubits. We present a technique using ion micromotion to reduce these sources of decoherence by over an order of magnitude. We benchmark the performance with a new method, based on randomized benchmarking, to estimate the magnitude of crosstalk errors on nearby qubits. Using the Honeywell System Model H\texttt{0}, we demonstrate measurement and reset on select qubits with low crosstalk errors on neighboring qubits.
\end{abstract}


\maketitle

\section{Introduction}
Measurement and reset on a subset of qubits in the middle of a quantum circuit is a fundamental operation for many quantum information protocols including teleportation~\cite{Bennett93}, error correction~\cite{Shor95,RyanAnderson21}, entanglement distillation~\cite{Bennett96}, measurement-based quantum computing~\cite{Briegel09}, and holographic simulation~\cite{FossFeig20,Chertkov21}. However, mid-circuit measurement and reset (MCMR) presents a system engineering challenge, requiring strong interactions with target qubits without inducing decoherence on nearby spectator qubits. Moreover, the irreversible nature of measurement and reset operations implies that unintended disturbances on spectator qubits (crosstalk errors) are incapable of being mitigated with standard techniques such as dynamical decoupling~\cite{Biercuk09}.

In trapped-ion systems, measurement and reset are performed by applying a resonant laser beam to the target qubit, causing the ion to fluoresce. There are two sources of MCMR crosstalk errors on neighboring qubit ions: 1) photons fluoresced by the measured ion and 2) photons originating from the measurement laser beam either directly or due to reflections from the trap or nearby surfaces. 

Previously, MCMR crosstalk in trapped-ion systems was reduced by three strategies. One strategy is to map the qubit to a secondary ion species where the detection can be performed at a laser frequency far away from the primary ion's resonant fluorescence frequency~\cite{Schmidt05,Hume07,Negnevitsky18}. This requires quantum information to be swapped between the different ion species, creating additional technical challenges, and imperfections in that process can limit the detection fidelity. However, errors due to imperfections in the mapping can be overcome by repeating the process as long as the qubit state information is preserved by the process as in ~\cite{Hume07} for example.  A second strategy is to map the qubit states of any unmeasured ions to different atomic levels that do not interact with the detection light~\cite{Riebe04}. This strategy requires additional laser pulses applied to all unmeasured qubits and can lead to additional qubit errors. A third strategy is to increase the distance between measured and unmeasured ions by physically transporting them to different trapping regions~\cite{Barrett04,Crain19,Pino21}. This strategy increases circuit time and may require the device to be larger than otherwise necessary. Furthermore, it has been found that even at separation distances of greater than $100$ $\mu$m, where decoherence due to the fluoresced photons of the measured ion is reduced to $10^{-4}$ levels or less, stray photons from the detection laser beam can be the limiting source of decoherence during measurements~\cite{Pino21,Crain19,Wan19}. 

Here, we present a method to ``hide'' ions using micromotion, which reduces unwanted photon absorption by over an order of magnitude~\cite{Berkeland98,Leibfried99}, and thereby also reducing the distance ions must be separated during measurement. We characterize the performance of the MCMR operations with a variant of randomized benchmarking (RB), where MCMR applied to a target qubit is interleaved between single-qubit RB gates on nearby qubits. The protocol simultaneously quantifies mid-circuit state preparation and measurement (SPAM) errors on the target ion and the crosstalk errors on the nearby ions.

This paper is organized as follows: In Sec.~\ref{sec:micromotion hiding} we introduce our technique to reduce MCMR crosstalk erors by hiding ions with micromotion. Next, in Sec.~\ref{sec:mcmr benchmarking} we introduce our procedure to comprehensively benchmark the crosstalk errors on neighboring ions. Finally, in Sec.~\ref{sec:conclusions} we discuss the results and future work.

\section{Micromotion hiding} \label{sec:micromotion hiding}
Our scheme to reduce MCMR crosstalk errors uses the RF field and DC electrodes present in Paul traps. In standard operation, RF fields and DC electrodes are usually used to trap ions at the RF null for gating and transport~\cite{Pino21}. However, here we intentionally displace ions from the RF field null by tuning the voltages of the DC trapping electrodes. Displacing the ion from the RF null causes periodic motion of the ion at the RF frequency $\Omega$, known as micromotion. If photons are impinging on the ion, micromotion along the propagation direction of the photon leads to a Doppler shift induced frequency modulation of the photons in the ion's rest frame~\cite{Berkeland98}.  By tuning the amount of micromotion, controlled by the magnitude of the ion's displacement from the RF null, it is possible to tune the frequency modulation index and perfectly suppress photon absorption at the carrier frequency. When the carrier transition is nulled, the absorption rate of photons will be limited by the frequency modulation sidebands, which are detuned from the carrier by integer multiples of the RF drive frequency $\Omega$. When $\Omega$ is larger than the atomic transition linewidth $\Gamma$, significant suppression of absorption can be achieved. 

Working in the pseudo-potential approximation, the amplitude of the micromotion is $A=\sqrt{2}\omega r/\Omega$, where $r$ is the ion's displacement from the null and $\omega$ is the radial secular frequency. The frequency modulation index is then given by $n=k A \cos(\theta)$ where $k$ is the wavenumber of the light and $\theta$ is the angle between the micromotion direction and the light propagation direction. The suppression of absorption, in the low light intensity limit (saturation parameter $\ll 1$), is then given by
\begin{equation}\label{eq:scat}
\frac{I(n)}{I_0} = J_0(n)^2+2\sum_{v=1}^{\infty}\frac{J_v(n)^2}{1+4v^2 \Omega^2/\Gamma^2},
\end{equation}
where $J_v(n)$ is the $v^{\text{th}}$ order Bessel function, $I_0$ is the scattering rate in the absence of micromotion, and $I(n)$ is the scattering rate with modulation index $n$. The maximal values of suppression occur when the first term, $J_0(n)$ is equal to $0$. 

\begin{figure} 
	\begin{center}
	  \includegraphics[width=\columnwidth]{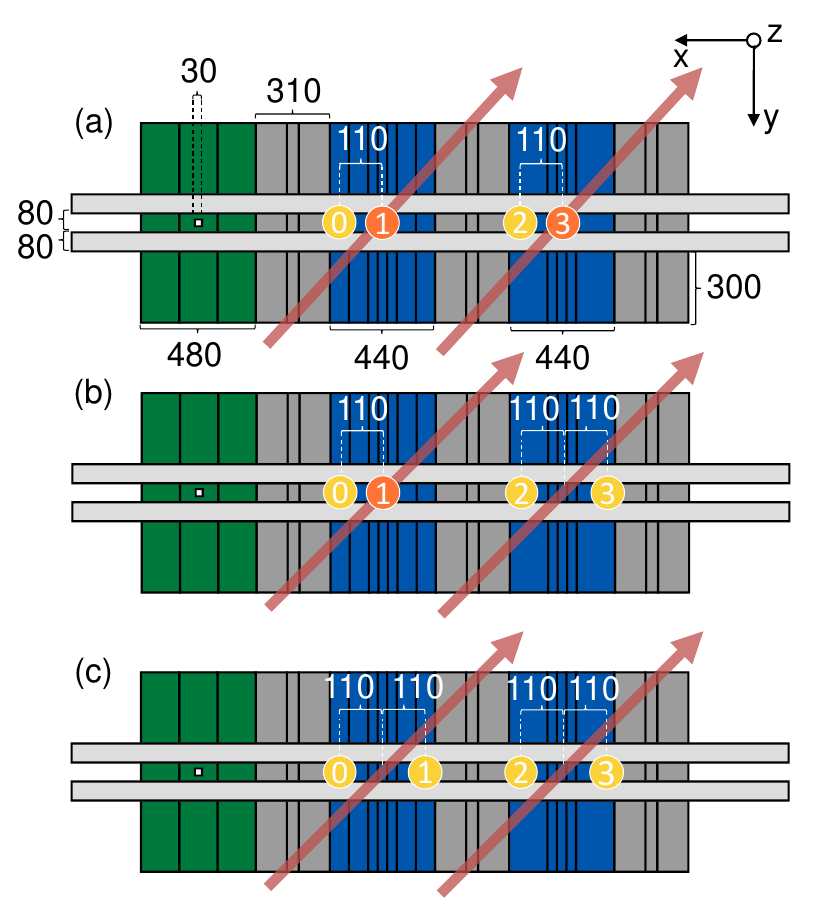}
	  \caption{Partial schematic of the ion trap presented in Ref.~\cite{Pino21}. All dimensions are given in $\mu$m. A load zone is shown on the left (green) with a through hole where neutral atoms emitted from an oven can enter. The two gate zones are colored blue, with gate zone 1 on the left with 7 DC electrodes segments and gate zone 2 on the right with 5 DC electrode segments, each spanning $440$ $\mu$m. Additional auxiliary zones consisting of three segments each are adjacent to the operation and loading zones. The locations of the ions are shown as circles along the center of the trap (each circle represents a Yb-Ba pair). Detection laser beams are shown as arrows crossing the trap at 45$^{\circ}$ to the trap axis. Measured ions are colored orange while unmeasured adjacent ions are colored yellow. (a) Parallel detection of ions in both gate zones with the closest unmeasured ion displaced $110$ $\mu$m to either the left or right side. (b) Detection of a single ion in gate zone 1.  The measurement beams cannot be switched independently, so in gate zone 2, two ions are each held $110$ $\mu$m away from the laser beam. (c) Detection of no ions in either operation zone.}
	\label{Trap}
	\end{center}
\end{figure}

We demonstrated the micromotion hiding technique in the Honeywell System Model H\texttt{0}, which is described in Ref.~\cite{Pino21}.  In brief, the system supports up to six $^{171}$Yb$^+$ ions, which encode qubits in the $S_{1/2}$ ground states ``bright'' $|1\rangle = |F=1,m_F=0\rangle$ and ``dark'' $|0\rangle = |F=0,m_F=0\rangle$, named because only the bright state will fluoresce photons during the measurement. Each $^{171}$Yb$^+$ ion is co-trapped with a $^{138}$Ba$^+$ ion, which is used for sympathetic cooling. The system has two quantum operation zones (gate zones 1 and 2) capable of performing two-qubit gates on pairs of qubits as well as single-qubit gates, measurement, and reset of individual qubits. Additional trapping regions serve as memory zones, and the qubits can be arbitrarily rearranged using basic transport primitives of splitting and recombining of ion crystals, linear transport, and ion crystal rotations.

Basic measurement and reset protocols are detailed in Ref.~\cite{Olmschenk07}. The measurement light is resonant with the transitions from the $S_{1/2}$, $F=1$ manifold to the  $P_{1/2}$ $|F=0,m_F=0\rangle$ excited state, which forms an approximately closed set of transitions and can only decay to $S_{1/2}$  $|F=0,m_F=0\rangle$ via off resonant coupling to the $P_{1/2}$, $F=1$ manifold, which is detuned by $2.1$ GHz. The $S_{1/2}$  $|F=0,m_F=0\rangle$ state can only absorb photons via transitions to the excited $P_{1/2}$, $F=1$ states, which are detuned by $14.7$ GHz from the measurement light and are thus highly suppressed, allowing measurement to distinguish between the two qubit states with high fidelity. Similarly, state intialization to $S_{1/2}$ $|F=0,m_F=0\rangle$ is achieved by applying a $2.1$ GHz sideband on the measurement beam using an electro-optic modulator. This additional sideband couples the $F=1$ ground state to the $P_{1/2}$ $F=1$ excited state, which can decay to $S_{1/2}|F=0,m_F=0\rangle$. This final state ceases to absorb photons due to the $14.7$ GHz detuning. 

During measurement and reset operations, a single  Yb$^+$-Ba$^+$ crystal is held in the center of a gate zone, while other ions are held at distant locations with the closest ion held 110 $\mu$m away to either the left or right side of the zone (Fig.~\ref{Trap}). The measurement duration is typically 120 $\mu$s, and approximately $900$ photons are scattered, of which we collect and measure with the detection system on average $1.7\%$, or about $15$ photons. These quantities are estimated by comparing the peak count rate observed with a model of the $^{171}$Yb$^+$ scattering rate. At a distance of 110 $\mu$m and ignoring possible reflections from the trap, the expected absorption probability for the unmeasured ion is estimated as $\approx 1.1\times 10^{-7}$ per photon fluoresced from the measured ion, or around $1.0\times 10^{-4}$ in total. State preparation and reset operations use the same laser beams as measurement and have a duration of 10 $\mu$s, which proportionally reduces the number of fluoresced photons and expected scattering rate.

The laser beams are designed to have a beam waist ($1/e^2$ radius) of 17.5 $\mu$m at the ion position and propagate parallel to the surface at an angle of 45$^{\circ}$ to the axis of the trap where all the zones are located. Assuming a Gaussian beam profile, the expected intensity at the position of the nearest ion would be negligible. However, departures from the ideal profile and scattering off of optics and surfaces leads to significant unwanted light at the unmeasured ions' positions, and this is the dominant source of MCMR error here. 

Static potentials that hold ions at various locations along the trap axis, including any transverse displacement from the rf null, are created by solving for a set of electrode voltages that satisfy the required electric field constraints at the desired ion locations, as described in~\cite{Pino21}. In order to induce micromotion along the direction of the laser beam, ion positions are displaced in the $\hat{y}$ direction shown in Fig.~\ref{Trap}. The actual locations of the ion positions are accurate up to imperfections in the electrostatic models used to generate the voltages and any uncompensated stray electric fields, which we estimate could lead to offsets between the actual and desired ion displacements from the rf null of at most 100 nm.  

We generated potentials with displacements from the RF null up to 5.5 $\mu$m, allowing us to explore suppression corresponding to the first two zeros of the $J_0$ function in Eq.~\ref{eq:scat}. In principle, the asymptotic form of the Bessel functions in Eq.~\ref{eq:scat} suggests that the absorption can be suppressed arbitrarily, scaling like $n^{-1}$. Electrode voltage solutions for larger displacements (up to 20 $\mu$m) can be found; however, the potential starts to deviate more from the desired solution as the displacement gets larger, and the voltages required start to approach the maximum amount allowed in our system. 

One possible concern in using the micromotion hiding technique is the potential heating of the ions' motion due to RF noise during displacement~\cite{Blakestad09}. We checked for additional heating by performing typical transport sequences to rearrange the ion ordering both with and without ion displacements at locations where a measurement might occur. We found that the final ion crystal mode temperatures after both types of sequences were indistinguishable, within the measurement error, of approximately one motional quanta. Furthermore, we found that in sequences including the displacements, the ions were still able to be recooled to the same final motional temperatures before two-qubit gates without any changes to the sympathetic cooling sequence. This indicates that any additional heating resulting from the ion displacements were negligible and would not adversely affect the execution of quantum circuits.

\section{Crosstalk characterization} \label{sec:mcmr benchmarking}
Here, we present two methods to characterize the suppression of photon scattering crosstalk errors. The first method uses a model of the depumping process out of the bright state to estimate the scattering rate. We use this procedure to verify the micromotion hiding technique is working and study the scattering rate as a function of ion displacement. The second method uses a new variation of RB to more fully characterize the photon scattering crosstalk errors. This second method requires fewer assumptions and quantifies different types of the errors, which could be used to estimate the effect on quantum circuits.

\begin{figure} 
	\begin{center}
		\includegraphics[width=\columnwidth]{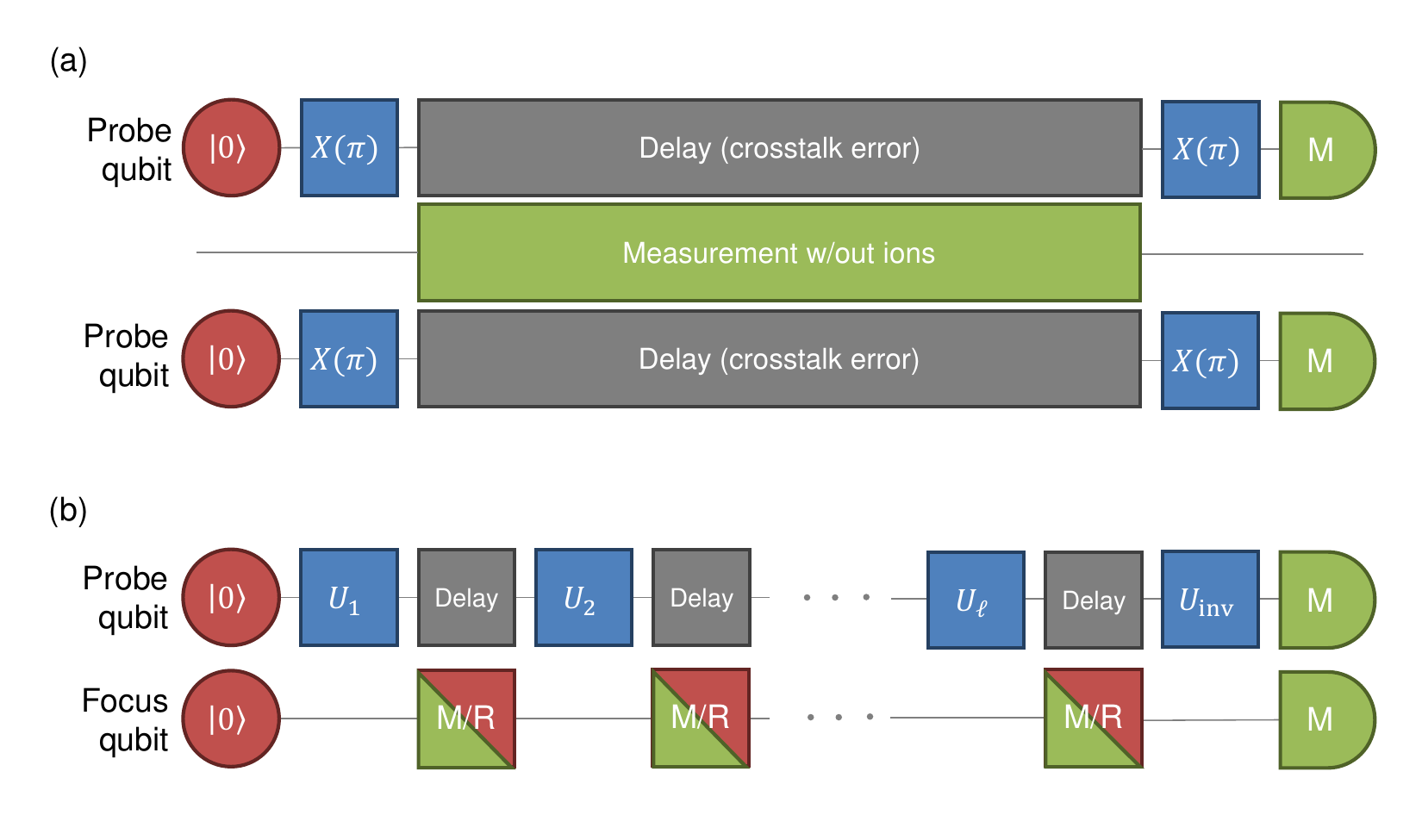}
		\caption{Circuit diagram for MCMR benchmarking procedures. Blue boxes show gates applied to qubits. Red shapes show state preparation. Green shapes show measurement. Grey boxes show delay time on qubits where crosstalk errrors occur due to nearby state preparations or measurement. (a) The bright state depumping method first prepares all ions in the bright state, then applies measurement pulses for a variable time. The decay in bright state population is proportional to the scattering rate under model assumptions described in the text. (b) The MCMR benchmarking procedure which applies RB pulses to nearby probe qubits interleaved with measurement and/or reset on focus qubits. Leakage and computational errors are extracted by fitting the decay of ideal measurement outcomes on the probe qubits under different assumptions about the structure of the error.}
		\label{fig:methods}
	\end{center}
\end{figure}

\subsection{Bright state depumping}
Our first method to quantify crosstalk errors measures depumping rates out of the bright state induced by photon scattering as illustrated in Fig.~\ref{fig:methods}a. For this technique, four ion qubits are prepared in the bright state $|1\rangle$ and shifted 110 $\mu$m to either side of the two gate zones as shown in Fig.~\ref{Trap}c. A first measurement pulse is then applied to both zones for a variable duration, but since all ions have been moved out of the center of the gate zone, no ions are actually measured.  Each qubit is then sequentially transported back to the center of the gate zones where we apply a spin-flip transition ($X(\pi)$ pulse) and then read out the state with a second measurement pulse of standard duration. Ideally, this leaves each qubit in the dark state $|0\rangle$ before the second measurement; so if no depumping occurred, we expect no photons to be detected (up to dark counts). If an ion scatters one photon during the first measurement pulse, then it will only be returned to the dark state $|0\rangle$ in approximately $1/3$ of the experiments. From careful analysis of the Lindblad master equation under the assumption of unpolarized light, we find that the decay of the bright state population is
\begin{equation} \label{eq:depump_rate}
p_{\textrm{depump.}}(t) = \tfrac{2}{3}\left(1 - e^{-3 \gamma t}\right),
\end{equation}
where $\gamma$ is the scattering rate and $t$ is the duration of the first measurement pulse. Derivation of the model and possible reasons for deviation of the asymptote from the ideal 2/3 value are discussed in Appendix~\ref{app:model}. To extract $\gamma$, we fit the decay in the bright state population as a function of $t$. Since no qubits are held in the gate zones during the first measurement pulse, depumping errors will only be caused by stray light from the measurement beam and not fluorescence from any measured ions. We expect only the former to benefit from the micrmotion hiding technique in the H\texttt{0} geometry. 

\begin{figure} 
	\begin{center}
	\includegraphics[width=\columnwidth]{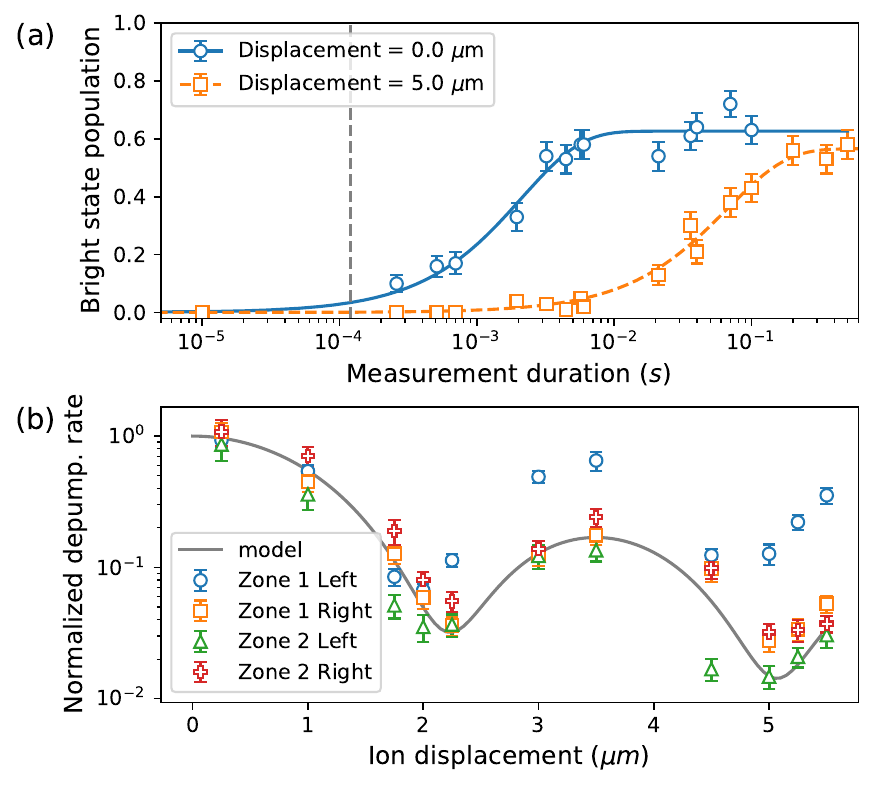}
	\caption{Bright state depumping data. (a) Depumping out of bright state induced by the measurement beam for an ion held 110 $\mu$m to the right of gate zone 2 when the ion is not displaced from the RF null (blue circles) and when it is displaced $5.0$ $\mu$m (orange squares). Each depuming curve is fit to an exponential decay shown as lines. The fitted characteristic times $(1/\gamma)$ are 6.4(8) ms and 199(17) ms respectively. The induced micromotion from the displacement significantly suppresses the depumping rate. Error bars represent the standard deviation of each measurement. The vertical dashed line is placed at the H\texttt{0} system measurement duration of 120 $\mu$s. (b) Depumping rates versus ion displacement from the RF null normalized to the depumping rate at zero displacement for an ion shifted $110$ $\mu$m to the left and right of each of the gate zones. Error bars represent the uncertainty in the fit parameters to Eq.~\ref{eq:depump_rate}. The solid line is derived from Eq.~\ref{eq:scat} and discussed in the main text. The characteristic time ($1/\gamma$) found from the fit to an exponential decay for each of the positions at zero displacement are: $18(1)$ ms, $1.0(1)$ ms, $1.6(3)$ ms, and $6.4(8)$ ms for zone 1 left, zone 1 right, zone 2 left, and zone 2 right respectively.}
	\label{fig:H0Data}
	\end{center}
\end{figure}

The results from the bright state depumping experiments are shown in Fig.~\ref{fig:H0Data}.  Fig.~\ref{fig:H0Data}a shows the depumped population versus measurement time for an ion in the gate zone 2 right position with zero and $5.0$ $\mu$m of displacement from the RF null, along with fits to the exponential decay model in Eq.~\ref{eq:depump_rate}. As can be seen, the induced micromotion from the displacement significantly suppresses the depumping.  Fig.~\ref{fig:H0Data}b shows the depumping rate for each ion position versus displacement from the RF null normalized to the depumping rates at zero displacement (values given in the figure caption).  The depumping rates at zero displacement span from $1.0(1)$ ms  to $18(1)$ ms. We found that this variation of scattered light was explained by steep fringes in the scattered light pattern that were strongly dependent on the particular beam alignment. The precise amplitude and location of the fringes were typically not stable on the multi-day time scale due to drifts in beam alignment.  The model shown with the data is based on Eq.~\ref{eq:scat} and includes no free parameters. The significant departure of the zone 1 left position data from the model is consistent with the presence of scattered light at that position propagating in a direction with an orthogonal component to the ion's micromotion, which limits the maximal suppression achieved. Nevertheless, in all cases, greater than 10$\times$ suppression in the depuming rate is achieved, demonstrating the effectiveness of the technique.

\subsection{MCMR benchmarking}
\begin{table*}[]
	\begin{tabular}{lcccc}
		\hline \hline
		Experiment                          & Focus qubits & Probe qubits & Initial state & Interleaved operation                                \\ \hline
		Control                       & -                               & $\{$0, 2$\}$                             & $\ket{0}$             & -                                                                        \\ \hline
		Reset            & $\{$1$\}$                               & $\{$0, 2$\}$                            & $\ket{0}$           & $\{$reset$\}$                                                                \\
		Dark measurement            & $\{$1$\}$                               & $\{$0, 2$\}$                            & $\ket{0}$           & $\{$measure$\}$                                                                \\
		Bright measurement              & $\{$1$\}$                               & $\{$0, 2$\}$                            & $\ket{1}$             & $\{$measure$\}$                                                                \\
		Dark measurement \& reset (1) & $\{$1, 3$\}$                            & $\{$0, 2$\}$                            & $\ket{0}$             & $\{$ measure, reset$\}$                                \\ \hline
		Dark measurement \& reset (2) & $\{$1, 3$\}$                            & $\{$0, 2$\}$                           & $\ket{0}$           & $\{$ measure, reset$\}$                               \\
		Bright measurement \& reset   & $\{$1, 3$\}$                             & $\{$0, 2$\}$                           & $\ket{1}$             & $\{$measure, reset, $X(\pi)\}$                     \\
		Bleed-through                 & $\{$1, 3$\}$                            & $\{$0, 2$\}$                             & $\ket{0}$             & $\{$random SU(2),measure, reset, measure, reset$\}$ \\ \hline \hline
	\end{tabular}
	\caption{MCMR benchmarking experiments. Each test is defined by the different focus qubits and probe qubits, initialization method $init = |0/1\rangle$, and interleaved operation. Qubit indexes follow the labeling convention in Fig.~\ref{Trap}.}
	\label{table:rb_def}
\end{table*}

Our second method is a variant of RB used to identify measurement and/or reset-induced crosstalk errors using fewer model assumptions. In this technique, we divide the available ions that contain qubits into focus qubits, which receive measurements and/or resets, and probe qubits, which suffer from measurement crosstalk. We apply standard single-qubit RB sequences to probe qubits interleaved with measurement and/or reset on focus qubits as illustrated in Fig.~\ref{fig:methods}b. The procedure then characterizes the crosstalk errors on probe qubits and mid-circuit SPAM errors on focus qubits. For these experiments we use a displacement of $r=2.3$ $\mu$m for all unmeasured ions, corresponding to the first zero of the $J_0$ function in Eq.~\ref{eq:scat}. We used the configuration in Fig.~\ref{Trap}a where ions one and three are probe ions and the configuration in Fig.~\ref{Trap}b where ion one is the only probe ion.

We first briefly describe standard RB, but further details can be found in Refs.~\cite{Magesan11, Baldwin19}. The standard RB procedure is to prepare an initial state (the dark state in this case), apply a sequence of random gates selected from a group (the Clifford group in this case), apply a final gate equivalent to the inversion of all previous gates and a random Pauli gate, and finally measure the overlap between the resulting state and the ideal final state defined by the random Pauli. We refer to this measured overlap as the survival probability $p$. Gate errors cause the survival probability to decrease from one scaling with the length of sequence and magnitude of errors. The standard RB analysis procedure is to measure the survival probability for several different, randomly generated circuits of various lengths $\ell$. Based on properties of the Clifford group and assumptions about the uniformity of errors for each gate, the average survival probability at a given length $\overline{p}(\ell)$ decays exponentially with rate $r$, which is proportional to the average fidelity of the gates~\cite{Magesan11, Wallman18}. The measured average survival probability is then fit to an exponential decay $\overline{p}(\ell) = A r^{\ell} + 1/2$ to extract the average fidelity.

We make use of two extensions to standard RB: leakage RB~\cite{Wood18} and interleaved RB~\cite{Magesan12}. Leakage RB is used when ``leakage errors'' connect the computational space to an extra subspace. Measurement and reset crosstalk are both examples of leakage errors. Leakage RB returns the rate that errors force population to leave (leakage $L$) or return (seepage $S$) to the computational subspace, defined as
\begin{equation}
\begin{split}
L = \tfrac{1}{2} \textrm{Tr}(\Lambda[\mathds{1}_c] \mathds{1}_e) \nonumber \\
S = \tfrac{1}{2} \textrm{Tr}(\Lambda[\mathds{1}_e] \mathds{1}_c)
\end{split}
\end{equation} 
where $\Lambda[\cdot]$ is the error channel in question, $\mathds{1}_c$ is the identity on the computational subspace, and $\mathds{1}_e$ is the identity on the extra subspace. Leakage RB analysis can be applied to the same dataset used for standard RB if the proper measurement outcomes are saved. The leakage analysis requires outcomes that project on to the computational subspace. This is like measuring the population left in the computational subspace, which in our case is the measured dark state population $p_0$.  The original leakage RB proposal~\cite{Wood18} uses assumptions about the gates to show that the average dark state population over a set of given length circuits $\overline{p}_0(\ell)$ decays exponentially with rate $t$ proportional to the leakage and seepage rates, but such assumptions require control that is unavailable in our system. Instead, we make reasonable assumptions about the errors being benchmarked; mainly that they are incoherent and symmetrically connect the extra subspace. Further details are in Appendix~\ref{app:benchmarking_methods}. With these assumptions, $\overline{p}_0$ decays exponentially as $\overline{p}_0(\ell) = A t^{\ell+1} + B$ with rate $t= 1 - L - S$.

\begin{table*}[]
	\small
	\begin{center}
		\begin{tabular}{lccccc}
			\hline \hline
			\multicolumn{2}{l}{}   & \multicolumn{2}{c}{Zone 1}           & \multicolumn{2}{c}{Zone 2}            \\ \hline
			Experiment & Focus qubits & Avg. error ($\times 10^{-3}$) & SPAM errors ($\times 10^{-3}$) & Avg. error ($\times 10^{-3}$) & SPAM errors ($\times 10^{-3}$) \\ \hline
			Control       & -   &  $0.19(6)$  & -   				 &  $0.31(1)$ & -    \\ \hline
			Reset         & $\{$1$\}$   & $0.23(6)$ & -    				 &   $0.34(7)$ & -    \\ 
			Dark measurement          & $\{$1$\}$   & $0.6(1)$ & $14.8(2)$    &   $1.3(1)$ & -    \\
			Bright measurement          & $\{$1$\}$   & $0.8(1)$ & $227.8(7)$$^*$    &   $1.6(2)$ & -    \\ 
			Dark measurement \& reset (1)      & $\{$1, 3$\}$   & $1.1(2)$ & $3.7(1)$    &   $2.0(2)$ & $2.28(8)$    \\ \hline
			Dark measurement \& reset (2)  & $\{$1, 3$\}$   & $0.54(9)$ & $1.45(6)$    &   $4.1(3)$ & $1.03(5)$   \\
			Bright measurement \& reset     & $\{$1, 3$\}$   & $0.46(9)$ & $5.0(1)$    &   $2.4(2)$ & $12.4(2)$    \\
			Bleed-through$^{\dagger}$     & $\{$1, 3$\}$   & $0.9(1)$ & $1.29(6)$    &   $6.3(5)$ & $1.19(6)$    \\\hline \hline
		\end{tabular}
		\caption{Estimated errors from MCMR benchmarking. $^*$This error rate is higher since any population projected to the dark state is likely to stay dark since there are no mid-circuit resets. $^\dagger$The SPAM error for the bleed-through experiment is the error from the second measurement.}
		\label{table:rb_data}
	\end{center}
\end{table*}

Interleaved RB uses the same sequences as standard RB but performs an additional experiment where a gate in question is interleaved between each randomized gate. The original interleaved RB proposal bounds the fidelity of the interleaved gate by comparing decay rates with and without the interleaved gate. In our case, we do not perform such an analysis and just measure the decay of the interleaved experiment. 

Our technique uses standard single-qubit RB gates on the probe qubits interleaved with mid-circuit measurement and/or reset on focus qubits as illustrated in Fig.~\ref{fig:methods}b. The interleaved measurement and/or reset ideally act as the identity on the probe qubits, but in practice cause crosstalk errors. The crosstalk error is then characterized by the computational space decay rate (measured in the standard analysis) and leakage/seepage rates (measured in the leakage analysis) from the same dataset. The mid-circuit operation is characterized by the SPAM errors when we apply both reset and measurement to the focus qubits.

To demonstrate the general procedure described above, we performed a series of MCMR benchmarking experiments designed to characterize errors in both operation zones caused by measurement and reset beam scattering, and by neighboring ion fluorescence. The different experiments' definitions are shown in Table~\ref{table:rb_def}. The tests were performed on two different days (day one consisted of the first five and day two the next three). The \textit{Dark measurement \& reset} test was performed twice to compare the rates on the different days.

\begin{figure} 
	\begin{center}
		\includegraphics[width=\columnwidth]{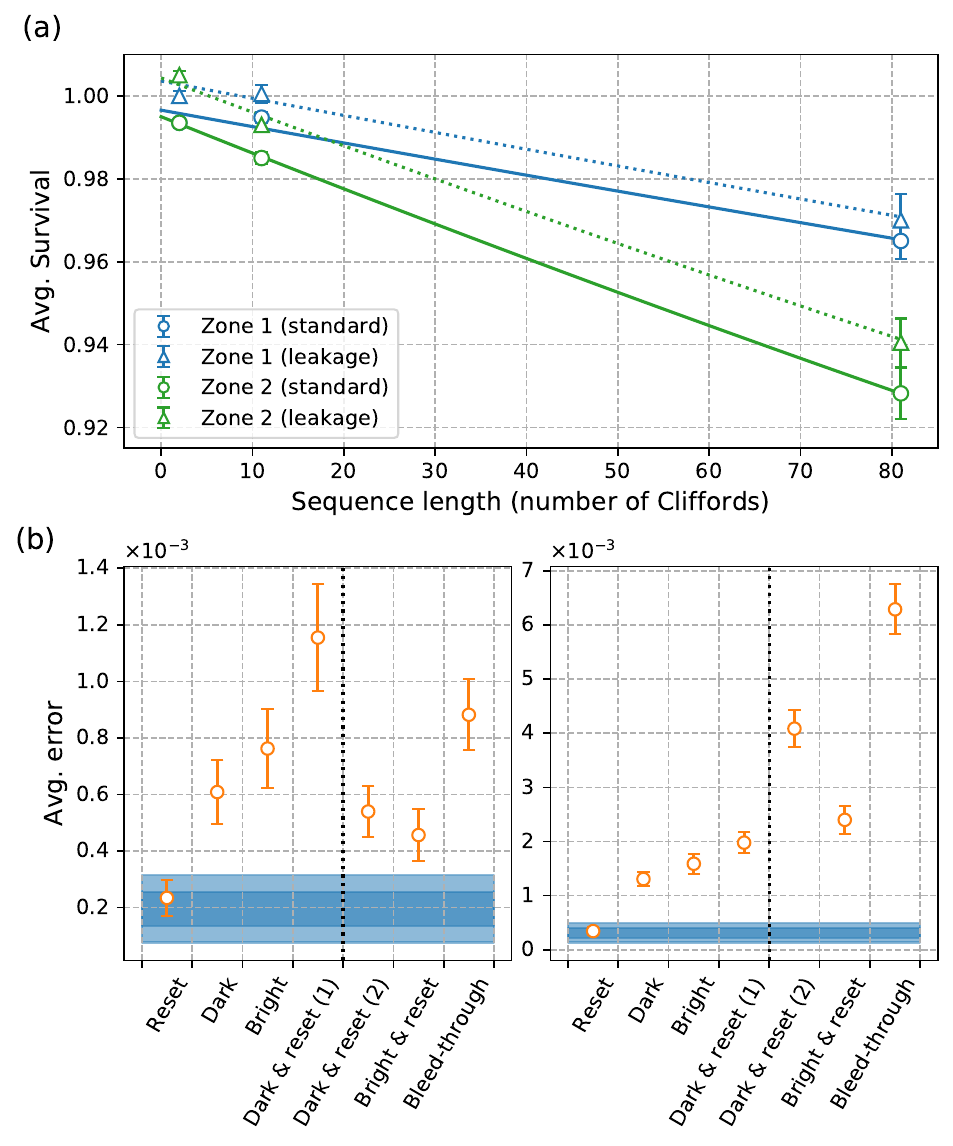}
		\caption{MCMR benchmarking results (a) Standard decay (solid lines with circles, blue for zone 1 and green for zone 2) and Leakage decay (dashed lines with triangles, blue for zone 1 and green for zone 2). Lines represent exponential fits to the data. Some points in the leakage analysis are above unit probability due to finite sampling and measurement errors. The leakage analysis points are found by averaging the dark outcomes over all sequences so are ideally 0.5 without errors but we normalize to one to match the standard analysis. (b) Estimated average error from MCMR benchmarking for zone 1 (left side) and zone 2 (right side) with various operations. Dark and light blue regions represent one- and two-sigma confidence interval for standard single-qubit RB (control test with no interleaved operations). Interleaved operations are labeled in correspondence with Table~\ref{table:rb_data}. The dashed black line divides data that was run on different days.}
		\label{fig:benchmarking}
	\end{center}
\end{figure}

For all MCMR benchmarking experiments we used 40 random sequences of length (2, 11, 81), and each sequence was repeated 100 times. An example MCMR decay curve is plotted in Fig.~\ref{fig:benchmarking}a. Our reported metrics are given below:
\begin{itemize}
\item \textit{Average error:}  Errors that occur on the computational subspace of the probe qubits are likely dominated by crosstalk errors from mid-circuit measurement and/or reset on the focus qubits. In this case the standard equation for average errors also includes leakage and is related to the standard decay and leakage rate, $r$ and $L$, by $\epsilon_{q} = (1-r + L)/2$. \\
\item \textit{Mid-circuit SPAM errors:}  This is the mid-circuit SPAM error rate for dark state or bright state preparation.
\end{itemize}
Uncertainties for the estimated RB quantities are reported as the standard deviation of distributions calculated from a semi-parametric bootstrapping method~\cite{Meier06}. Uncertainties in the SPAM errors are calculated from a binomial distribution standard deviation based on the number of total shots. 

The results are given in Table~\ref{table:rb_data} and also plotted compared to the control experiment in Fig.~\ref{fig:benchmarking}b. The micromotion hiding technique yields MCMR crosstalk errors $<5 \times 10^{-3}$ in both zones but zone 1 shows better performance than zone 2. It also shows that the mid-circuit SPAM errors are consistent with end of circuit errors and (as discussed later) are independent of circuit location.

In the first three tests, zone 1 has both probe and focus qubits while zone 2 only has two probe qubits. This allows us to compare the MCMR crosstalk errors on probe qubits with and without neighboring focus qubits that receive measurement and/or reset. Even though both ions are displaced outside of the center of zone 2 we still measured crosstalk errors in both zones due to scattering in the system. In fact, we found crosstalk errors are slightly worse in zone 2, which was also seen with the bright state depumping measurements. Reset operations cause smaller errors than measurement since the resonant light is applied for a shorter duration. Reset is also on resonance with a different hyperfine sublevel (discussed in Appendix~\ref{app:model}) causing a lower ratio of leakage-to-seepage, and therefore a shallower decay. The reset errors are consistent with the estimated error from single-qubit RB and likely smaller than the precision set by the choice of sequence length, number of randomizations, and shots. We expect more photons to be fluoresced from the focus qubits for the bright measurement test than the dark measurement test. However, the measured error was roughly consistent up to uncertainty between the two indicating, again, that ion florescence is not the leading contribution to the crosstalk error.

In the next four tests, each zone contained a focus and probe qubit and also measures the mid-circuit SPAM errors and reset bleed-through. These experiments (except for \textit{Dark measurement \& reset (1)}) were performed at a later date. Comparing the \textit{Dark measurement \& reset tests 1 \& 2}, which were run on each day, it appears day 2 had slightly larger MCMR crosstalk errors in zone 2 but smaller than zone 1, possibly due to slight changes in the optical modes. The final \textit{Bleed-through} experiment, designed to quantify reset bleed-through, contains two resets, two measurements and a single qubits gates, and therefore had a larger crosstalk error than other experiments. Interestingly, the \textit{Dark measurement \& reset (2)} test had larger errors than the \textit{Bright measurement \& reset} test in zone 2. This may be due to some drift in the system.

\begin{figure} 
	\begin{center}
	  \includegraphics[width=\columnwidth]{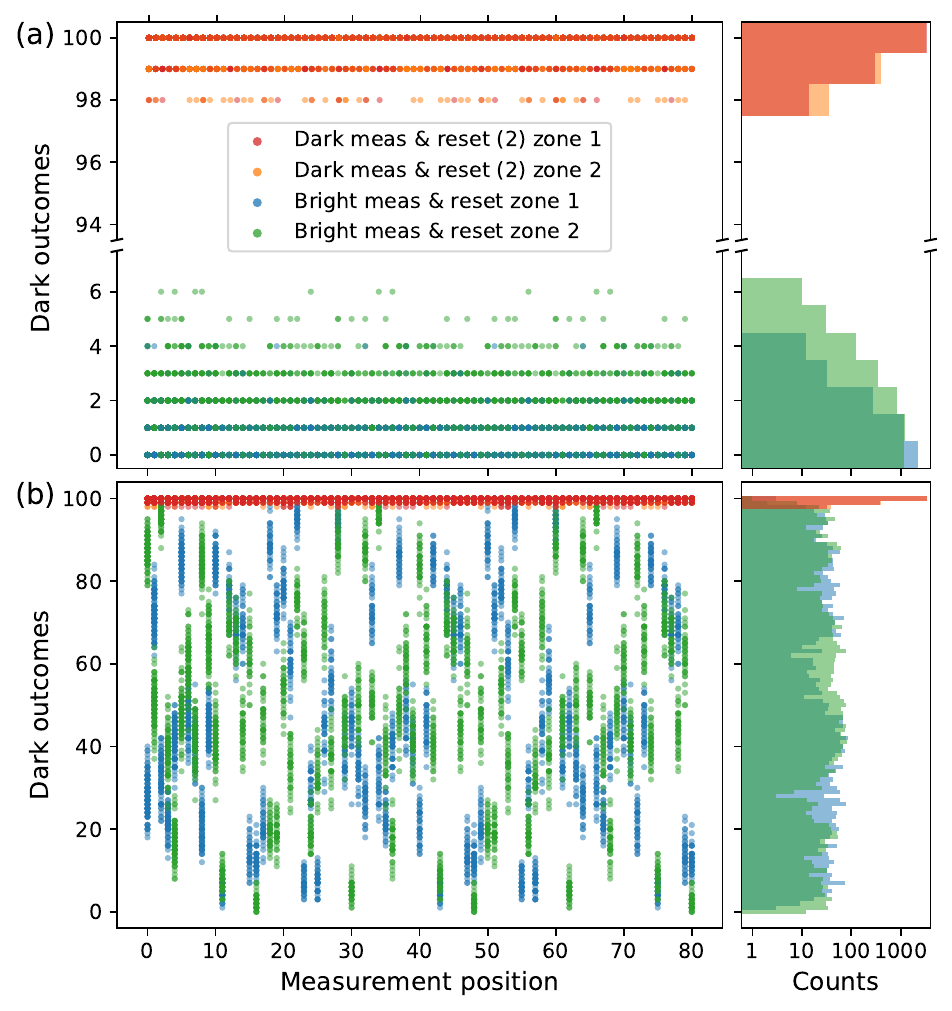}
	  \caption{Experimental data from MCMR benchmarking SPAM operations. (a) Data collected in \textit{Dark measurement \& reset (2)} and \textit{Bright measurement \& reset} datasets, which prepare dark and bright states respectively on focus qubits interleaved between every Clifford on probe qubits. Orange points and bars are qubit 1 and red points and bars are qubit 3 in \textit{Dark measurement \& reset (2)}. Blue points and bars are qubit 1 and green points and bars are qubit 3 in \textit{Bright measurement \& reset}. (b) Data collected in \textit{Bleed-through} dataset. Blue points and bars are qubit 1 and green points and bars are qubit 3 correspond to the first measurement and orange points and bars are qubit 1 and red points and bars are qubit 3 from the second measurement.}
	\label{fig:spam}
	\end{center}
\end{figure}

The measured mid-circuit SPAM errors were estimated from \textit{Dark measurement \& reset (2)} and \textit{Bright measurement \& reset} and are plotted in Fig.~\ref{fig:spam}. Notably, the number of dark states detected appears independent of the circuit length indicating that mid-circuit SPAM errors are constant for the entire circuit. The average mid-circuit SPAM error was $3.2(1)\times 10^{-3}$ for zone 1 and $6.7(1) \times 10^{-3}$ for zone 2. During these tests the end-of-circuit SPAM error was also measured to be $2(1)\times 10^{-3}$ in zone 1 and $7(1)\times10^{-3}$ in zone 2 (via standard single-qubit RB). Therefore, end-of-circuit SPAM error and mid-circuit SPAM errors were within two standard deviations for zone 1 and one standard deviation for zone 2. 

The \textit{Bleed-through} experiment demonstrates that a single reset is effective at reinitializing the state as shown in Fig.~\ref{fig:spam}b. The random initial gate leaves the focus qubits in a superposition of dark and bright states as verified by the first measurement. The reset then destroys any residual information about the population as verified by the second measurement with SPAM error of $1.29(6)\times 10^{-3}$ in zone 1 and $1.19(6)\times 10^{-3}$ in zone 2, which is within two-sigma of the average mid-circuit SPAM error from the \textit{Dark measurement \& reset} test.

Finally, we perform an analysis to see how consistent our experimental measurements from bright state depumping and MCMR benchmarking are with the model of measurement crosstalk. We derived the photon scattering from the different measured parameters (bright state depumping, standard decay, and leakage decay) under the extra assumption that the photon scattering is small. The relations and comparisons between results are shown in Table~\ref{table:comparison}. The estimates from each method are similar in each zone although some fall outside one standard deviation. The leakage analysis appears to return lower estimates of the scattering rate than the other methods, which is likely due to shallower decays in the leakage data. Other discrepancies may be due to some uneven mixture of polarization that is not accounted in the model but we study numerically in Appendix~\ref{app:numerics}.

\begin{table}[]
\small
\begin{center}
\begin{tabular}{lccc}
\hline \hline
Experiment & Relation & Zone 1 & Zone 2 \\ \hline
Bright state depumping                 & - &  $0.75(5)$  & $2.7(3)$ \\ \hline
Dark meas. \& reset (2)    & $3(1 - r)/4$   & $0.6(1)$  &  $4.0(3)$  \\ 
Dark meas. \& reset (2)    & $2(1 - t)/3$   & $0.3(1)$  &  $3.8(5)$  \\  \hline
Bright meas. \& reset       & $3(1 - r)/4$  & $0.51(9)$  & $2.6(2)$ \\ 
Bright meas. \& reset       & $2(1 - t)/3$   & $0.3(1)$ &  $1.8(4)$  \\  \hline \hline
\end{tabular}
  \caption{Estimated scattering rate ($\times 10^{-3}$) from different experimental methods in standard measurement time (120 $\mu$s). The estimate for the bright state depumping measurement is for the left ion configuration at 2.3 $\mu$m to match the benchmarking experiment. Uncertainty from the bright state depumping measurement is taken from covariance of the exponential fit. Uncertainty from the benchmarking estimates is taken from semi-parametric bootstrap uncertainty estimate.}
\label{table:comparison}
\end{center}
\end{table}

\section{Conclusions} \label{sec:conclusions}
We have demonstrated a new protocol to reduce mid-circuit measurement and reset crosstalk errors by hiding ions with micromotion. The protocol's effectiveness is verified using a new method based on RB, and results showed that mid-circuit measurement and reset can be performed with high fidelity on a subset of targeted qubits. The error rates reported in this work are approaching reported estimates for thresholds required in quantum error correction protocols~\cite{Tomita14}. The crosstalk error rates can be improved further with specially designed electrode geometries and improved optical mode quality. Our methods help to alleviate significant challenges to implementing important algorithms in transport based trapped-ion quantum computer architectures.

\begin{acknowledgments}
	We would like to thank the entire Honeywell Quantum Solutions team for contributions to the System Model H\texttt{0}. We especially thank the Honeywell System Model H1 team for helpful discussion and Alex Hall for operating the System Model H\texttt{0} experiment.
\end{acknowledgments}

\appendix
\section{Overview of system} \label{app:model}
\begin{figure} 
	\begin{center}
		\includegraphics[width=1.5\columnwidth]{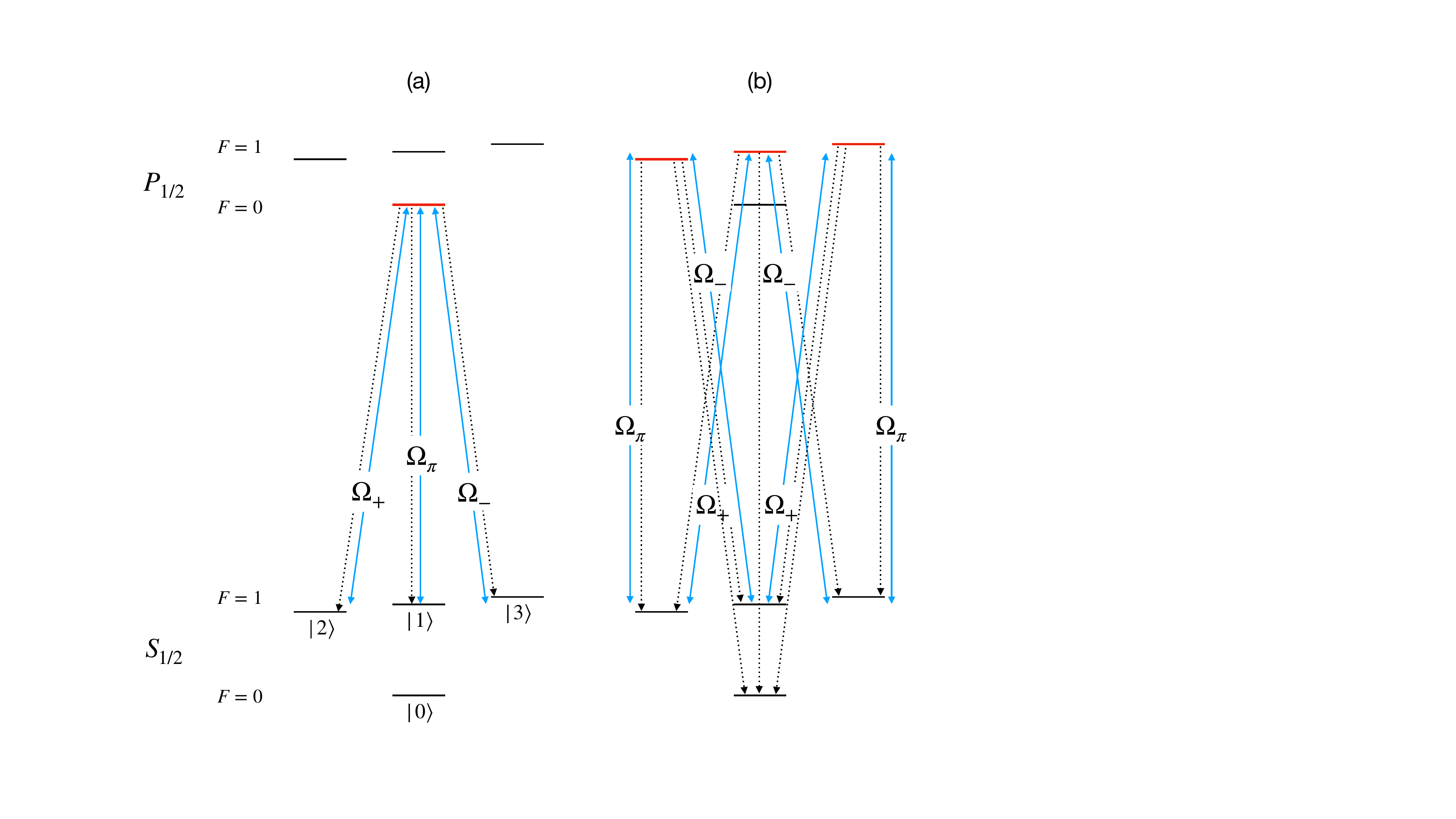}
		\caption{Illustration of atomic structure and laser couplings relevant for (a) measurement and (b) reset.}
		\label{fig:structure}
	\end{center}
\end{figure}

In our system, the computational space $\mathcal{H}_c$ is spanned by the basis states $|F=0,m_F=0\rangle=|0\rangle$ and $|F=1,m_F=0\rangle=|1\rangle$ of the $S_{1/2}$ electronic ground state of $^{171}$Yb$^{+}$, which are approximate hyperfine clock states at small magnetic fields. MCMR errors cause population to move into an extra subspace $\mathcal{H}_e$ spanned by the basis states $|F=1,m_F=-1\rangle =|2\rangle$ and $|F=1,m_F=+1\rangle =|3\rangle$ as shown in Fig.~\ref{fig:structure}. Single-qubit gates act on the computational subspace but still cause some rotation on the extra subspace due to small AC Stark shifts. The gates are ideally separable into a unitary on the computational subspace and a unitary on the extra subspace: $U = V_c \oplus W_e$.  

We model both measurement and reset by writing down the full Hamiltonian involving laser coupling between $S_{1/2}$ and $P_{1/2}$ and integrating out all vacuum modes for emitted photons to derive a master equation in the Born-Markov approximation (both measurement and reset also involve repumping ions that have decayed to $D_{3/2}$ via the $D[3/2]_{1/2}$ state, but the low branching ratio from $P_{1/2}$ to $D_{3/2}$ for Yb makes the repump process a small perturbation on the otherwise closed $S_{1/2}\leftrightarrow P_{1/2}$ transition).  Since measurement and reset beam intensities are designed to saturate the $S_{1/2}\rightarrow P_{1/2}$ transition for target ions, it is safe to assume that probe ions see Rabi rates $\Omega_{+,-,\pi}\ll\Gamma$, with $\Gamma$ the $P_{1/2}$ line width, and we can therefore adiabatically eliminate the $P_{1/2}$ states to derive a master equation governing the dynamics amongst states $\ket{\alpha}$ ($\alpha=0,1,2,3$).  Moreover, the effective scattering rates between states in the $S_{1/2}$ manifold, which scale as $\Omega^2/\Gamma$, are generally much less than the energy splittings between the various states, and so we can also adiabatically eliminate all ground-state coherences to derive a classical rate equation describing pumping of populations $P_{\alpha}$ between the various states $\ket{\alpha}$,
\begin{align}
\label{eq:rate_equation}
\dot{P}_{\alpha}=-P_{\alpha}\sum_{\beta}R_{\alpha\rightarrow\beta}+\sum_{\beta}R_{\beta\rightarrow\alpha}P_{\beta}.
\end{align}
Measurement is performed by coupling $S_{1/2},\,F=1$ to $P_{1/2},\,F=0$.  Decay from $P_{1/2},\,F=0$ down to $S_{1/2},\,F=0$ is forbidden, leading (up to the repump caveat described above) to a cycling transition, and therefore the measurement dynamics takes place within the subspace $\alpha=1,2,3$, with the population of the $\ket{0}$ state constant.  

In steady state (the relevant situation for target qubits) and assuming one only measures total fluorescence and is agnostic to the distribution of population within the $F=1$ manifold, the measurement process is well described by projections into $F=1$ and $F = 0$ submanifolds via POVM elements $E_0 = |0 \rangle \langle 0|$ and $E_1 = |2 \rangle \langle 2| + | 1 \rangle \langle 1| + |3 \rangle \langle 3|$. To quantify measurement induced dynamics of probe qubits in the transient regime, and to properly account for the distributions of populations within $F=1$, we instead resort to Eq.\,(\ref{eq:rate_equation}), finding rates
\begin{align}
R_{\alpha\rightarrow\beta}=\frac{\Omega^2_{\alpha}\Gamma}{\Gamma^2/4+\Delta_{\alpha}^2}.
\end{align}
Here $\Delta_{\alpha}$ are the Zeeman shifts of states $\ket{1,2,3}$, and the rates are independent of final state due to equality of all Clebsch-Gordan coefficients. Measurement beam polarizations are calibrated to maximize the total scattering rate on target ions, which occurs when transition rates are equal out of each state. Denoting this constant rate $R_{\alpha\rightarrow\beta}=\gamma$, the solution to the rate equation with boundary condition $P_{\alpha}(0)=\delta_{\alpha,1}$ (i.e., starting in the qubit $\ket{1}$ state) is
\begin{align}
P_1(t)&=\frac{2}{3}\left(e^{-3\gamma t}+\frac{1}{2}\right)\\
P_2(t)&=P_3(t)=\frac{1}{3}(1-e^{-3\gamma t}).
\end{align}
The bright state depumping experiment described in the paper (results in Fig.\,\ref{fig:H0Data}) measures $P_2+P_3$, giving rise to Eq.\,(\ref{eq:depump_rate}) in the manuscript.  In Fig.\,\ref{fig:H0Data} we treat the amplitude of $P_2(t)+P_3(t)$ (nominally $2/3$) as a fit parameter, and see small deviations from $2/3$; this deviation likely arises from imperfect measurement beam calibration, or differential scattering of the different polarizations off surfaces.

\section{Benchmarking methods} \label{app:benchmarking_methods}
In this Appendix, we detail our method for benchmarking MCMR based on leakage and interleaved benchmarking. One important note is that this method relies on assumptions about the form of the errors present, motivated by previous diagnostics and understanding of the system. This is different than the standard RB paradigm that makes minimal assumptions about errors, but we believe it is appropriate in this setting. We outline the system, assumptions made, and derive the decay rates used to extract error information.

\subsection{Representation background}
In the following, we use the Liouville representation, which is useful in dealing with general error channels and determining the average behavior over a set of circuits. The Liouville representation translates operators on the Hilbert space, such as density operators or POVM elements, to supervectors $A \rightarrow | A \rrangle$. The inner product of two supervectors is equal to the trace overlap between the corresponding operators $\llangle A | B \rrangle =\textrm{Tr}(A^{\dagger} B)$ (i.e. the Born Rule). The Liouville representation also translates completely positive processes that act on operators, in general described by a Kraus decomposition, to superoperators $\sum_i A_i \rho A_i^{\dagger} \rightarrow \sum_i A_i^* \otimes A_i | \rho \rrangle = \mathcal{A} | \rho \rrangle$. A trace preserving (TP) process maps the adjoint identity supervector to itself $\llangle \mathds{1} | \mathcal{A} = \llangle \mathds{1} |$.

Define a basis of orthonormal operators (and corresponding supervectors) for the four-dimensional space, $\textrm{Tr}(P_i^{\dagger} P_j) = \llangle P _i | P_j \rrangle = \delta_{i,j}$ Four elements are a basis for operators on $\mathcal{H}_c$: $\{\mathds{1}_c \oplus 0_e, X_c \oplus 0_e, Y_c \oplus 0_e, Z_c \oplus 0_e\}$, which are the standard Pauli operators. Likewise, four other elements are similarly defined on $\mathcal{H}_e$, and the remaining 8 elements are Pauli-XY-like operators that describe coherences between the two subspaces, which in operator form is $X_{ce,i,j} = |i_c \rangle\langle j_e| + |j_e \rangle \langle i_c|$ and $Y_{ce,ij} = -i |i_c \rangle\langle j_e| + i|j_e \rangle \langle i_c|$.

Ideal gates are broken up into unitary rotations on different operator subspaces, and in the Liouville representation,
\begin{equation} \label{eq:exp_uni}
\mathcal{U} =  (V_c \oplus W_e)^* \otimes (V_c \oplus W_e) =\mathcal{V} + \mathcal{W} + \mathcal{A},
\end{equation}
where $\mathcal{V} = (V_c \oplus 0_e)^* \otimes (V_c \oplus 0_e)$ spans the $\mathcal{H}_c$ operators, $\mathcal{W} = (0_c \oplus W_e)^* \otimes (0_c \oplus W_e)$ spans the $\mathcal{H}_e$ operators, and $\mathcal{A} = (V_c \oplus 0_e)^* \otimes (0_c \oplus W_e) + (0_c \oplus W_e)^* \otimes (V_c \oplus 0_e)$ spans the Pauli-XY-like operators. The superoperator $\mathcal{U}$ is block diagonal in the operator subspaces introduced above ($\mathcal{V} \mathcal{W} = \mathcal{V} \mathcal{A} = \mathcal{A}\mathcal{W} = 0$).  In future subsections we drop subscript $c$ and $e$ since the subspace should be clear from other labels and context. 

\subsection{Randomized benchmarking decay}
In this subsection we provide a derivation for the MCMR benchmarking decay. An MCMR benchmarking experiment proceeds as follows:
\begin{enumerate}
\item Prepare each qubit in the dark state $\ket{0}$, 
\item Apply a random single-qubit gate selected from the Clifford group to probe qubits,
\item Apply a measurement and/or reset to focus qubits,
\item Repeat steps 2-3 $\ell$ times,
\item Apply a final inversion gate to the probe qubits compiled to undo all previous gates and include an additional random Pauli gate, 
\item Measure either the population in the dark submanifold ($F=0$ when random Pauli = $\mathds{1}$ or $Z$) or bright submanifold ($F=1$ when random Pauli = $X$ or $Y$),
\item Repeat steps 1-6 $s$ times,
\item Repeat steps 1-7 for different values of $\ell$,
\item Fit resulting measurement record according to standard and leakage analyses defined below.
\end{enumerate}
One difference between this and most standard RB schemes is including a random Pauli gate compiled into the final inversion gate, which allows us to differentiate the leakage error rates from the computational error rates~\cite{Baldwin19,Harper19}. Importantly, we pick the same number of random $\mathds{1}$ and $Z$ Paulis as random $X$ and $Y$ Paulis. This is critical for the stability of the leakage analysis and explained in more detail later.

The general procedure to derive the RB analysis is to calculate the survival probability for an individual sequence and then use it to determine the average survival probability over all sequences of a fixed length. The second part requires a few assumptions about the error channel, mainly that it is constant across all gates. In our analysis, we add additional assumptions to deal with leakage errors.

For an individual MCMR benchmarking sequence, let $i$ index the sequence, $j$ index the final Pauli compiled into the inversion, and $k$ index the measurement (dark $k=0$ or bright $k=1$). The survival probability for such an indexed sequence is
\begin{equation} \label{eq:RB_dist}
p_{i,j,k}(\ell) = \llangle k | \Lambda_M \Lambda_{i,j,\textrm{inv}} \mathcal{U}_{i, j, \textrm{inv}}\Lambda_{i,\ell}\mathcal{U}_{i,\ell} \cdots \Lambda_{i,1} \mathcal{U}_{i,1} \Lambda_P |0 \rrangle,
\end{equation}
where $\Lambda_P$ and $\Lambda_M$ are SPAM errors and $\Lambda_{i,m}$ is the combination of single-qubit gate errors and the measurement/reset crosstalk error (for gate position indexed by $m$). These errors may have any form and couple the computational and extra subspaces.

Next, we specify the assumptions we use to determine the average survival probability with leakage errors. Ref.~\cite{Wood18} makes four assumptions about the \textit{gates} that are applied. These assumptions all require a control of the extra subspace and relative phase between the two subspaces, which is not available in our system. Instead, for our procedure we make three assumptions about the \textit{errors} for measurement/reset crosstalk that lead to the same conclusions. Usually, one avoids making assumptions about the errors in RB experiments; however, in this specialized situation where we have a good description of the dominant error channel, it is justified. Our assumptions are:
\begin{itemize}
\item \textit{Assumption E1}: Measurement crosstalk error dominate single-qubit gate errors. From our bright-state depumping measurement and standard RB we find that the mid-circuit measurement crosstalk error is about an order of magnitude larger than single-qubit gate errors. Therefore, the error channel from the interleaved measurement dominate. Our procedure is also constructed to have the same operation after each Clifford gate assuming Markovian errors. (This is equivalent to the first assumption in Ref.~\cite{Wood18}.)
\item \textit{Assumption E2}: SPAM errors only cause symmetric and incoherent leakage/seepage errors. For state preparation, any population that leaks into the extra subspace is evenly distributed between the two states and not coherent with the population in the computational subspace. For measurement, the POVM elements are diagonal in the z-basis (i.e. measurement axis is aligned with the quantization axis) and the extra states are measured evenly.
\item \textit{Assumption E3}: Measurement and reset crosstalk errors are incoherent and symmetric. Same as the assumption above, the crosstalk errors evenly distribute between each state in the extra subspace and do not maintain any coherence with population left in the computational subspace.
\end{itemize}
Assumption E1 does not hold for mid-circuit reset operations but we find that the procedure still works well in simulations as shown in Appendix~\ref{app:numerics}.

Assumptions E2 and E3 are strong assumptions about the errors in the system that are justified from the analysis in Appendix~\ref{app:model} and previous measurements of the system. First, measurement and reset are driven by spontaneous emission processes and, therefore, inherently incoherent processes by design. Second, any coherence that may survive could be eliminated by varying the relative time between operations which randomizes over coherent states and cancels any residual terms. Finally, in our system, any asymmetry between the basis elements of the extra subspace results from uneven light polarizations. We believe this is unlikely from other diagnostics but could also be eliminated by applying RF fields after every gate to randomize the population. 

The net result of Assumptions E1 and E3 is that we can assume the errors from MCMR have a fixed form for all sequences $i$ and gate locations $m$,
\begin{equation} \label{eq:lambda}
\Lambda = \Lambda_c + \Lambda_e + \sum_{i \in \{0, 3\}} \lambda^{c \rightarrow e}_{i} |\mathds{1}_{e} \rrangle \llangle P_{c,i}| +  \lambda^{e \rightarrow c}_{i} |P_{c,i} \rrangle \llangle \mathds{1}_{e}|,
\end{equation}
where $\Lambda_c$ and $\Lambda_e$ only contain errors on each respective subspace and $\lambda^{c\rightarrow e}_{i}$ and $\lambda^{e\rightarrow c}_{i}$ are the rates at which population moves between the subspaces with $i \in \{0, 3\}$ for the Paulis $\mathds{1}_{c}$ and  $Z_{c}$. This is roughly the same form that Ref.~\cite{Wood18} forced by their four assumptions on the gates. 

Returning to Eq.~\eqref{eq:RB_dist}, we apply Assumption E1 and expand each $\mathcal{U}_{i,m}$ in terms of the action on each subspace,
\begin{widetext}
\begin{equation}
p_{i, j, k}(\ell) = \llangle k | \Lambda_M \mathcal{P}_j (\mathcal{V} + \mathcal{W} + \mathcal{A})_{i, \textrm{inv}}\Lambda (\mathcal{V} + \mathcal{W} + \mathcal{A})_{i,\ell} \cdots \Lambda (\mathcal{V} + \mathcal{W} + \mathcal{A})_{i,1} \Lambda_P |0 \rrangle,
\end{equation}
\end{widetext}
where we set $\Lambda_{i,j,\textrm{inv}} = \mathds{1}$ by simply not applying measurement/reset after the final inversion gate and we have separated out the final random $\mathcal{P}_j$, which we assume acts as the identity on operator basis elements outside the computational subspace.

Based on Assumption E2, the initial state (with preparation errors) only spans the computational subspace operators and $|\mathds{1}_e\rrangle$. Moreover, by assumptions E3 there are no other terms that can populate other operator basis elements. Therefore, we can treat the action of $\mathcal{A}$ as identity since it will not change the state. The small $Z$ rotations on the extra subspace also do not change the state's component in the extra subspace since there is no coherence in that subspace. Therefore, we can also treat the action of $\mathcal{W}$ as identity.

Now, we are ready to derive the average survival probability over all sequences; that is an average over all possible gates in each position for fixed sequence length, final random Pauli, and measurement. As in standard RB~\cite{Magesan11}, we expand each Clifford gate acting on the computational subspace $\mathcal{U}_{i,m} = \mathcal{D}_{i,m}^{\dagger} \mathcal{D}_{i, m -1}$ where $\mathcal{D}_i$ is a relabeling of the representations of Clifford group elements,
\begin{equation} \label{eq:pavg}
\begin{split}
\overline{p}_{j, k}(\ell) &= \tfrac{1}{N_C^{\ell}} \sum_i \llangle k | \Lambda_M  
\mathcal{P}_j  \mathcal{U}_{i,\textrm{inv}} \Lambda \mathcal{U}_{i,\ell} \, \cdots \,
\Lambda \mathcal{U}_{i,1}  \Lambda_P
|0 \rrangle,  \\
&= \llangle k | \Lambda_M  \mathcal{P}_j \left[ \tfrac{1}{N_C} \sum_i  \mathcal{D}_{i}  \Lambda  \mathcal{D}_{i}^{\dagger}\right]^{\ell}
\Lambda_P |0 \rrangle.
\end{split}
\end{equation}

In standard RB, without an extra subspace, the summation inside the brackets in Eq.~\eqref{eq:pavg} is referred to as a twirl. The twirl projects any error to a purely depolarizing error channel~\cite{Magesan11}. However, with the extra subspace the action is more complicated. We substitute Eq.~\eqref{eq:lambda} from Assumptions E3,
\begin{align} ~\label{eq:Lambda_T}
\Lambda_T &= \tfrac{1}{N_C} \sum_i 
\mathcal{D}_{i}(  \Lambda_c +\sum_{i} \lambda^{c \rightarrow e}_{i} |\mathds{1}_{e} \rrangle \llangle P_{c,i}| +  \lambda^{e \rightarrow c}_{i} |P_{c,i} \rrangle \llangle \mathds{1}_e|) \mathcal{D}_{i}^{\dagger}, \nonumber \\
&= r \mathbb{P}_c + t_c | \mathds{1}_c \rrangle \llangle \mathds{1}_c | \nonumber \\
& + L | \mathds{1}_e \rrangle \llangle \mathds{1}_c | + S| \mathds{1}_c \rrangle \llangle \mathds{1}_e | + t_e |
\mathds{1}_e \rrangle \llangle \mathds{1}_e |
\end{align}
where $\mathbb{P}_c$ is the projection onto the part of the RB subspace orthogonal to $| \mathds{1}_c \rrangle \llangle \mathds{1}_c|$. The coefficients are defined $r = \textrm{Tr}(\mathbb{P}_c \Lambda)$, $L =\llangle \mathds{1}_e|\Lambda |\mathds{1}_c \rrangle$, $S = \llangle
\mathds{1}_e| \Lambda |\mathds{1}_c \rrangle$, $t_c=\llangle \mathds{1}_c| \Lambda |\mathds{1}_c
\rrangle=1-L$, $t_e =\llangle \mathds{1}_e| \Lambda |\mathds{1}_e \rrangle = 1- S$.

Next, we diagnolize Eq.~\eqref{eq:Lambda_T} to solve for $\Lambda_T^{\ell}$. The first term in Eq.~\eqref{eq:Lambda_T} commutes with all other terms $\mathbb{P}_1 |\mathds{1}_{c/e} \rrangle \llangle \mathds{1}_{c/e}| = 0$, but the other terms span a two-dimensional subspace $\{ | \mathds{1}_c \rrangle, | \mathds{1}_e \rrangle\}$ decomposed as
\begin{equation}
\begin{pmatrix} \label{eq:tp_mat}
1- L & S \\
L & 1-S
\end{pmatrix} 
= t_+ \Pi_+ + t_- \Pi_-,
\end{equation}
with eigenvalues $t_{\pm}$ and eigenvectors $\Pi_{\pm}$,
\begin{align} \label{eq:pi_pm}
t_+ &= 1, \nonumber \\
t_- &= 1-L - S, \nonumber \\
\Pi_+ &= \frac{1}{ L +  S}
\begin{pmatrix} 
 S & S\\
 L &  L 
\end{pmatrix}, \nonumber \\
\Pi_- &= \frac{1}{L + S}
\begin{pmatrix} 
 L &  -  S\\
-L &  S
\end{pmatrix}.
\end{align}
Then $\Lambda_T^{\ell} = r^{\ell} \mathbb{P}_c + \Pi_+ + t_-^{\ell} \Pi_-$. 

Substituting in these expressions we can derive the average survival probability for each outcome and final Pauli,
\begin{equation} \label{eq:p_k}
\overline{p}_{j, k}(\ell) =A_{j,k} r^{\ell} + B_{k} t_-^{\ell} + C_{k},
\end{equation}
were $A_{j,k} =  \llangle k| \Lambda_M \mathcal{P}_j \mathbb{P}_c \Lambda_P | 0 \rrangle$, $B_{k} =\llangle k| \Lambda_M \Pi_+ \Lambda_P | 0 \rrangle \\$ and $C_{k} =  \llangle k| \Lambda_M  \Pi_- \Lambda_P | 0 \rrangle$ since $\mathcal{P}_j \Pi_{\pm} = \Pi_{\pm}$.

We now derive the decay for each analysis by solving for $A_{j,k}$, $B_{j,k}$, and $C_{j,k}$ for different measurement outcomes:

\textit{Standard Analysis}: Choose $k=0$ when $j = 0,3$ and $k=1$ when $j=1,2$. This corresponds to selecting the measurement outcome with the largest expected population based on the final Pauli. Then
\begin{equation}
\begin{split}
p_S(\ell) &= \tfrac{1}{4}\left[ \bar{p}_{0, 0}(\ell) + \bar{p}_{1, 1}(\ell) + \bar{p}_{1, 2}(\ell) + \bar{p}_{0, 3}(\ell) \right], \nonumber \\
	&= \overline{A} r^{\ell} + \tfrac{1}{2},
\end{split}
\end{equation}
where $\overline{A} = \tfrac{1}{4}(A_{0, 0}+ A_{1, 1}+A_{1, 2}+A_{0, 3})$, $\sum_k B_k = 0$, and $\tfrac{1}{4}\sum_k C_k= \tfrac{1}{2}$. The reduction of $B$ and $C$ terms can be derived from definitions of $\mathcal{P}_j$, $\Pi_{\pm}$ and $E_0$ and $E_1$ given above.

\textit{Leakage Analysis}: Choose $k=0$ for all $j$. Then
\begin{equation} \label{eq:leakage_decay}
\begin{split}
p_L(\ell) &= \tfrac{1}{4}\sum_j \bar{p}_{0, k}(\ell), \\
	&= B_0 t_-^{\ell+1} + C_0.
\end{split}
\end{equation}
This reduction comes from the fact that $\sum_j \mathcal{P}_j = |\mathds{1}_c \rrangle \llangle \mathds{1}_c| + \mathds{P}_{\notin c}$ where $\mathds{P}_{\notin c}$ is the projection onto the rest of the operator space not including the computational subspace operators. The extra $t_-$ comes from  $\tfrac{1}{4}\sum_k B_k = B_0 t_-$. The coefficients $B_0$ and $C_0$ can be expanded in terms of the leakage and seepage operators assuming no leakage/seepage in SPAM
\begin{equation}
\begin{split}
B_0 &= \frac{L/2}{L + S}, \\
C_0 &= \frac{S/2}{L + S},
\end{split}
\end{equation}
which relates $L$ and $S$ to the ratio between the y-intercept and the asymptote $L/S = B_0/C_0$. Leakage/seepage SPAM produce additive errors with the same magnitude as the leakage/seepage SPAM rates as shown in Ref.~\cite{Wood18, Baldwin19}. Ignoring leakage/seepage SPAM, we can solve for $L$ and $S$ individually
\begin{equation}
\begin{split}
L &= 2 B_0 (1 - t_-), \\
S &= 2 C_0 (1 - t_-).
\end{split}
\end{equation}
In practice, we use $C_0$ to estimate $L$ as shown in Ref.~\cite{Wood18} since it has lower additive error from leakage/seepage SPAM.

The average error (average infidelity) over the computational space can also be calculated based on the twirled error channel,
\begin{equation}
\epsilon = 1 - f = \frac{1 - r + L}{2}.
\end{equation}

From Eq~\eqref{eq:leakage_decay} it is clear why it is beneficial to choose the number of $\mathds{1}/Z$ final Paulis to equal the number of $X/Y$ final Paulis. The condition forces $\textrm{Var}[p_L(\ell)] = 0$ without errors but over a finite sample of sequences. If we were to sample from the full set of Pauli gates, then $\textrm{Var}[p_L(\ell)] \geq 0$ even without errors, causing $p_L(\ell)$ to vary between different sequence length due solely to finite sampling and result in false identification of leakage and seepage errors.

For the derivation, we assumed that every possible sequence is measured, but in practice we only sample a finite set. For standard RB, the number of sequences required is reasonable for most experiments~\cite{Helsen19}. We have not verified our method follows similar scaling nor do we strictly comply with derived requirements. However, in the next section we perform simulations with the expected errors and show that there is good agreement between injected and estimated error rates with the number of sequences used. 

\subsection{Numerical simulations} \label{app:numerics}
Here, we present numerical simulations of MCMR benchmarking for both measurement and reset crosstalk errors. We also simulate uneven polarizations to see how the method responds.

\begin{figure} 
	\begin{center}
		\includegraphics[width=1.1\columnwidth]{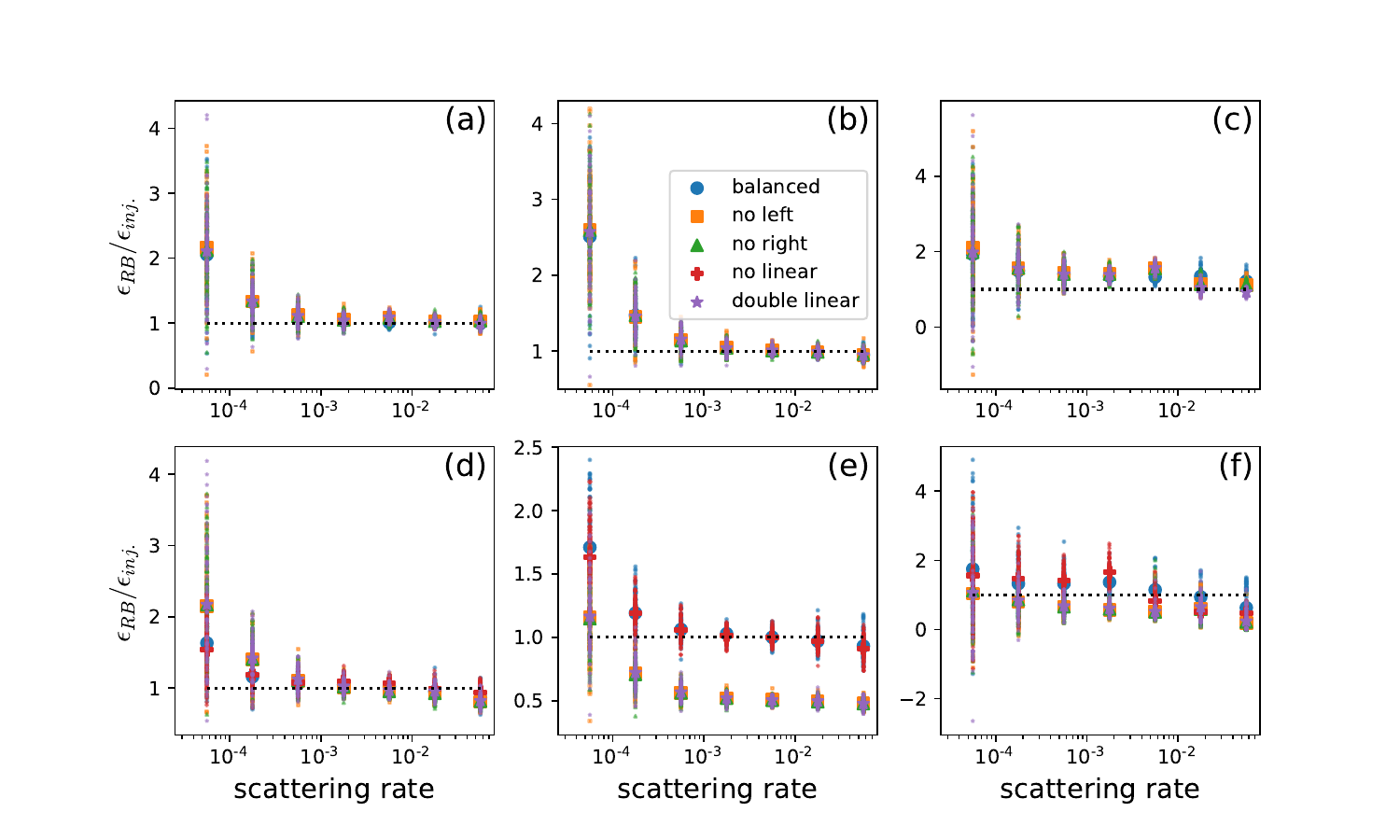}
		\caption{Simulation of MCMR benchmarking with ratio of estimated average error from RB to injected average error vs the injected scattering rate. Error models include: (Blue circles) balanced polarization, (orange squares) no left circular polarization, (green triangles) no right circular polarization, (red pluses) no linear polarization, (purple stars) linear polarization is double circular polarziation. Left column (a and d) results for average fidelity estimation. Center column (b and e) results from estimated scattering from standard decay analysis. Right column (c and f) results from estimated scattering from leakage analysis. Top measurement crosstalk errors. Bottom reset crosstalk errors.}
		\label{fig:numerics}
	\end{center}
\end{figure}

We generate several different error models by varying the error magnitude and polarization balance. For each error model, we apply MCMR benchmarking with 40 random sequences of length $\ell = (2, 11, 81)$ and repeat each with 100 shots. We perform a fitting to the measured survival probabilities with the methods outlined above for the standard and leakage analyses. We repeat the MCMR procedure with 100 different sets of random sequences and data to see the spread in performance. For each type of error, we initialize our fits based on the expected ratio of leakage to seepage and find that the fitting is sensitive to the initial guess, which is a drawback to the procedure, but the expected ratio is clear in most contexts.

For each error model, we calculate the average fidelity and scattering parameter from both the standard and leakage analyses as done in the main text. The estimate for the scattering parameter is done assuming even polarizations, which is violated by design in this test.

The simulations results are plotted in Fig.~\ref{fig:numerics}. A few trends are clear from the simulations. First, the method returns accurate estimates for average fidelity with either measurement or reset crosstalk errors with reasonable error magnitudes. The estimate of the scattering rate varies since even polarization is assumed in the derivation, which is violated in all but one error model. For example, the method underestimates the scattering rate for uneven polarization in reset crosstalk. This is due to the different relation between the standard and leakage decays and the scattering parameter. Second, the method performs best in the range (few)$\times 10^{-4}$ to (few)$\times 10^{-1}$ due to the choice of sequence lengths. To get better precision for smaller errors we could include longer sequences or more random sequences per length. Third, measurement crosstalk is easier to extract than reset crosstalk. This is because $L=S$ for measurement crosstalk but $L\leq S$ for reset crosstalk since reset pumps population to the dark state by design. This makes the decay in the leakage analysis much more shallow since $L/S = B_0/C_0$.

\end{document}